\documentclass[twocolumn,noshowpacs,preprintnumbers,amsmath,amssymb,aps,pra,groupedaddress,showpacs]{revtex4}
\usepackage[colorlinks,citecolor=blue,linkcolor=red,hyperindex,CJKbookmarks]{hyperref}

\usepackage{amsmath}
\usepackage{amsthm}
\usepackage{graphicx}
\usepackage{dcolumn}

\usepackage{mathrsfs}
\usepackage{amssymb}
\usepackage{amsfonts}

\usepackage{fancyhdr}
\begin{document}

\title{Experimental demonstration of entanglement-enabled universal quantum cloning in a circuit}

\author{Zhen-Biao Yang$^{1}$}
\author{Pei-Rong Han$^{1}$}
\author{Xin-Jie Huang$^{1}$}
\author{Wen Ning$^{1}$}
\author{Hekang Li$^{2}$}
\author{Kai Xu$^{2,3}$}
\email{kaixu@iphy.ac.cn}
\author{Dongning Zheng$^{2,3}$}
\author{Heng Fan$^{2,3}$}
\email{hfan@iphy.ac.cn}
\author{Shi-Biao Zheng$^{1}$}
\email{t96034@fzu.edu.cn}
\affiliation{1.Fujian Key Laboratory of Quantum Information and Quantum Optics, College of Physics and Information Engineering, Fuzhou University, Fuzhou, Fujian 350108, China}
\affiliation{2.Institute of Physics and Beijing National Laboratory for Condensed Matter Physics, Chinese Academy of Sciences, Beijing 100190, China}
\affiliation{3.CAS Center for Excellence in Topological Quantum Computation, University of Chinese Academy of Sciences, Beijing 100190, China}

\begin{abstract}
No-cloning theorem forbids perfect cloning of an unknown quantum state. A universal quantum cloning machine (UQCM), capable of producing two copies of any input qubit with the optimal fidelity, is of fundamental interest and has applications in quantum information processing. This is enabled by delicately tailored nonclassical correlations between the input qubit and the copying qubits, which distinguish the UQCM from a classical counterpart, but whose experimental demonstrations are still lacking. We here implement the UQCM in a superconducting circuit, and investigate these correlations. The measured entanglements well agree with our theoretical prediction that they are independent of the input state, and thus constitute a universal quantum behavior of the UQCM that was not previously revealed. Another feature of our experiment is realization of deterministic and individual cloning, in contrast to previously demonstrated UQCMs, which either were probabilistic or did not constitute true cloning of individual qubits.
\end{abstract}

\maketitle
\section{INTRODUCTION}
An unknown quantum state cannot be cloned perfectly due to the linearity associated with
the unitary transformation of quantum mechanics. This feature, discovered by Wooters and
Zurek in 1982 and known as the no-cloning theorem \cite{wootters_nature1982}, represents one of the fundamental
differences between quantum information and classical information. In particular, it ensures
the security of quantum cryptography schemes \cite{gisin_rmp2002,scarani_rmp2005,fan_pr2014}.

Because of the impossibility of perfect quantum cloning, much attention has
been paid to the possibility of producing copies close to the original
states. In the seminal paper by Buzek and Hillery, a universal quantum
cloning machine (UQCM) was proposed, which produces two identical
approximate copies via controllably entangling them with the original qubit
\cite{stackrel_pra1996}. The output state of each of these two copy qubits has a fidelity of 5/6
to the input state, which is independent of the input state and was proven
to be optimal \cite{gisin_prl1997,brubeta_pra1998}. Besides fundamental interest, quantum cloning can be
used to improve the performance of some quantum computational tasks \cite{galvao_pra2000}, to
distribute quantum information, and to realize minimal disturbance
measurements \cite{ricci_prl2005}.
The UQCM has been reported in nuclear magnetic resonance systems \cite{cummins_prl2002,du_prl2005}, but where the true cloning of
individual quantum systems cannot be achieved due to the ensemble aspect. Huang et al. presented a proof-of-principle demonstration
in an optical system \cite{huang_pra2001}, where only a single photon was involved; its polarization state was copied
onto one path freedom degree.

Several optical experiments have been reported, where
the state of a photon was copied onto another photon \cite{linares_science2002,fasel_prl2002,ricci_prl2004,irvine_prl2004,nagali_naturephotonics2009}, but the cloning processes are probabilistic for lack of a deterministic two-qubit controlled
gate between different photons in these experiments. 

Besides the limitation of ensemble aspect or probabilistic nature, previous experiments did not reveal the nonclassical correlations between the original input qubit and the copying qubits. These correlations enable the information carried by the input state to be equally imprinted on the clones with the optimal fidelity, and represent the most fundamental difference between the UQCM and a classical cloning machine. Quantitative characterization of these correlations is important for revealing the genuine quantum behavior of the UQCM, which is closely related to the universality and optimality of the copying operation.

We here adapt a scheme proposed in the context of cavity quantum electrodynamics [18] to a superconducting circuit involving Xmon qubits controllably coupled to a bus resonator. The high degree of control over the qubit-qubit interactions enables realizations of all gate operations required for approximately cloning the state of each qubit in a deterministic way. We indicate the universality of the implemented UQCM, and quantitatively characterize the entanglement between the input qubit and each of the copy qubits. The results confirm our theoretical prediction that this entanglement is also input-state-independent, and represents a universal quantum feature of the UQCM. 
The entanglement between the two copy qubits is also measured.

%

\section{RESULTS}
\subsection{Implementation of UQCM.}
\begin{figure*}[t]
	\centering
	\includegraphics [width=5in]{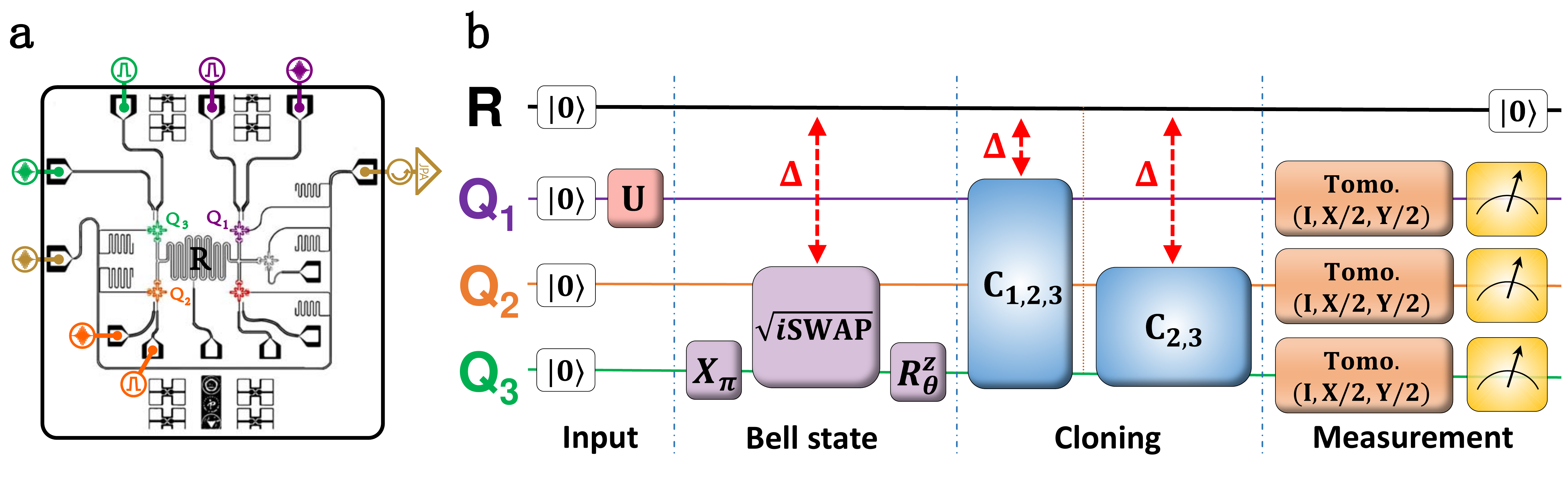}
	\caption{Device image and experimental scheme for the implementation of UQCM. \textbf{a} The device image. The device sample has five superconducting Xmon Qubits (three of them are used and are labeled from $Q_1$ to $Q_3$) and a central bus resonator $R$; the qubits' manipulation and their coupling to the resonator and the measurement are controlled by microwave pulses injected onto the circuit sample. \textbf{b} Before the copying operation, all qubits are initialized to their ground state at the corresponding idle
		frequencies. The whole procedure can be divided into four steps: Preparation
		of the input state through a unitary rotation at the idle frequency,
		denoted as $U$; entanglement of $Q_{2}$ and $Q_{3}$ with Bell type, achieved by a $\pi $
		rotation $X_{\pi }$ on $Q_{3}$, the $Q_{2}$-$Q_{3}$ $\sqrt{i\text{SWAP}}$
		gate, and a small Z pulse on $Q_{3}$ realizing a rotation $R_{\theta}^z$ for phase compensation; cloning of the
		input state onto $Q_{2}$ and $Q_{3}$, realized by resonator-induced
		couplings $C_{1,2,3}$ and $C_{2,3}$; output state tomography. $C_{1,2,3}$ is
		implemented by tuning $Q_{2}$ and $Q_{3}$ on resonance with $Q_{1}$ at the
		working frequency, while $C_{2,3}$ realized by tuning $Q_{1}$ back to its
		idle frequency, leaving $Q_{2}$ and $Q_{3}$ coupled to each other. Note that in our experiment, steps 1 and 2 are completed simultaneously for the sake of reducing qubits' decoherence, see Supplementary Note 4 for the details.}
	\label{f1}
\end{figure*}

The sample used to perform the experiment involves five Xmon qubits \cite{song_naturecommunications2017}, 
three of which are employed in our experiment and 
labeled from $Q_{1}$ to $Q_{3}$; these qubits are almost symmetrically coupled to a
central bus resonator, as sketched in Fig. \ref{f1}a. The resonator
has a fixed frequency of $\omega _{\text{r}}/2\pi=5.588$ GHz, while the frequencies of the
qubits are individually adjustable, which enables us to tailor the system
dynamics to accomplish the copying task. The Hamiltonian for the total
system is

\begin{equation}
H=\hbar \left[ \omega _{\text{r}}a^{\dagger }a+\sum_{j=1}^{3}\omega
_{\text{q},j}\left\vert 1_{j}\right\rangle \left\langle 1_{j}\right\vert
+\sum_{j=1}^{3}g_{j}\left( a^{\dagger }S_{j}^{-}+aS_{j}^{+}\right) \right] ,
\end{equation}

where $a^{\dagger }$ and $a$ are the photonic creation and annihilation
operators for the resonator, respectively, $S_{j}^{+}=\left\vert
1_{j}\right\rangle \left\langle 0_{j}\right\vert $ and $S_{j}^{-}=\left\vert
0_{j}\right\rangle \left\langle 1_{j}\right\vert $ are the flip operators
for $Q_{j}$, with $\left\vert 0_{j}\right\rangle $ and $\left\vert
1_{j}\right\rangle $ being its ground and first excited states separated by
an energy gap $\hbar \omega _{\text{q},j}$, $g_{j}$ are the corresponding
qubit-resonator coupling strengths, and $\hslash $ is the reduced Planck
constant. In our sample these coupling strengths are almost identical, e.g.,
$g_{j}\simeq g\simeq 2\pi\times$20 MHz. The system parameters are detailed in Supplementary 
Note 3. The qubit frequency tunability makes the system dynamics
programable.

When two or more qubits are detuned from the
resonator by the same amount much larger than $g$, they are coupled by
virtual photon exchange \cite{zheng_prl2000,zheng_prl2001,zheng_pra2010,osnaghi_prl2001,fedorov_nature2012,reed_nature2012,zhong_prl2016,song_prl2017,xu_prl2018,ning_prl2019,song_science2019}.
In our experiment, $Q_{1}$ acts as the
original qubit whose state is to be cloned, and $Q_{2}$ and $Q_{3}$ are used
as the copying qubits. 

The experimental sequence for realizing the UQCM with our setup is shown in
Fig. \ref{f1}b. The experiment starts with initializing the resonator to the vacuum
state $\left\vert 0_{r}\right\rangle $ and the qubits to their ground state $%
\left\vert 0_{1}0_{2}0_{3}\right\rangle $ at their idle frequencies. These
idle frequencies are highly detuned from the resonator frequency and
off-resonant with each other, ensuring each qubit to be effectively
decoupled from the resonator and other qubits when staying at its idle
frequency. After the initialization,
a suitable rotation is applied to $Q_1$ to prepare it in the state to be cloned
\begin{equation}
\left\vert \psi _{\text{in}}\right\rangle =\alpha \left\vert
0_{1}\right\rangle +\beta \left\vert 1_{1}\right\rangle ,
\end{equation}%
where $\alpha $ and $\beta $ are complex numbers, satisfying $\left\vert
\alpha \right\vert ^{2}+\left\vert \beta \right\vert ^{2}=1$. Prior to the
copying operation, we have to prepare $Q_{2}$ and $Q_{3}$ in the entangled
state $\left\vert \psi _{2,3}^{+}\right\rangle =(\left\vert
1_{2}0_{3}\right\rangle +\left\vert 0_{2}1_{3}\right\rangle )/\sqrt{2}$. To
prepare this state, we first transform $Q_{3}$ to the excited state $%
\left\vert 1_{3}\right\rangle $ by a $\pi $ rotation $X_{\pi }$, and then
tune $Q_{2}$ and $Q_{3}$ to the working frequency $\omega _{\text{w}}/2\pi= 5.44$ GHz.
With this setting, the resonator will not exchange photons with the
qubits and remain in the ground state due to the large detuning, but can
mediate energy swapping between the two qubits \cite{zheng_prl2000}.\quad After a duration
of $57.7$ ns, a $\sqrt{i\text{SWAP}}$ gate is realized, which evolves these two
qubits to the state $(e^{i\theta}\left\vert 1_{2}0_{3}\right\rangle +\left\vert
0_{2}1_{3}\right\rangle )/\sqrt{2}$, where $\theta=\pi/2+\theta_{\text{d}}$, with $\theta_{\text{d}}$ being
the extra dynamical phase accumulated during the frequency tuning process. To cancel the phase $\theta$,
$Q_{3}$ is tuned to the frequency $%
5.311$ GHz, where the rotation $R_{\theta}^z=e^{i\theta \left\vert 1_3\right\rangle\left\langle 1_3\right\vert}$ is realized after a duration of $30$ ns.

\begin{figure*}[t]
	\centering
	\includegraphics[width=7in]{Figure2_singlerho_v1-eps-converted-to.pdf}
	\caption{Reconstructed output states of $Q_{2}$ and $Q_{3}$ for
		six input states: \textbf{a} $\left\vert 0_{1}\right\rangle $; \textbf{b} $\left(
		\left\vert 0_{1}\right\rangle +i\left\vert 1_{1}\right\rangle \right) /\sqrt{%
			2}$;  \textbf{c} $\left(\left\vert 0_{1}\right\rangle -i\left\vert 1_{1}\right\rangle \right) /\sqrt{%
			2}$; \textbf{d} $\left( \left\vert 0_{1}\right\rangle +\left\vert 1_{1}\right\rangle
		\right) /\sqrt{2}$; \textbf{e} $\left( \left\vert 0_{1}\right\rangle -\left\vert 1_{1}\right\rangle
		\right) /\sqrt{2}$; \textbf{f} $\left\vert 1_{1}\right\rangle $. The measured output
		density matrices of $Q_{2}$ and $Q_{3}$ are respectively displayed in the
		upper and lower panels. Each matrix element is characterized by two
		color bars, one for the real part and the other for the imaginary part. The
		black wire frames denote the corresponding matrix elements of the output
		states yielded by the perfect UQCM. }
	\label{f2}
\end{figure*}
\bigskip After the production of $\left\vert \psi _{2,3}^{+}\right\rangle $,
$Q_{2}$ and $Q_{3}$ are tuned on resonance with $Q_{1}$ at the working
frequency, where these qubits are red-detuned from the resonator by the same
amount $\Delta=2\pi\times 148$ MHz. With this setting, the resonator does not exchange
photons with the qubits due to the large detuning, but can mediate a
coupling of strength $\lambda =g^{2}/\Delta $ between any two of these
qubits. The resonator will remain in the ground state during this process,
and can be discarded in the description of the system dynamics. In the
interaction picture, the state evolution of the qubits is governed by the
effective Hamiltonian \cite{zheng_prl2000,zheng_prl2001}

\begin{equation}
H_{\text{e}}=-\lambda \sum_{j,k=1}^{3}S_{j}^{+}S_{k}^{-},\text{ }j\neq k.
\end{equation}%
Under this Hamiltonian, $Q_{2}$ and $Q_{3}$ symmetrically interact with $%
Q_{1}$ through excitation exchange, with the number of the total excitations
being conserved. After an interaction time $\tau =2\pi /9\lambda $, the three-qubit coupling $C_{1,2,3}$ evolves $Q_1$, $Q_2$ and $Q_2$ to the entangled state
\begin{eqnarray}
&&\alpha \left[ \sqrt{\frac{2}{3}}\left\vert 1_{1}\right\rangle \left\vert
0_{2}\right\rangle \left\vert 0_{3}\right\rangle +\sqrt{\frac{1}{3}}e^{-i\pi
/3}\left\vert 0_{1}\right\rangle \left\vert \psi _{2,3}^{+}\right\rangle %
\right] \\
&&+\beta \left[ \sqrt{\frac{2}{3}}\left\vert 0_{1}\right\rangle \left\vert
1_{2}\right\rangle \left\vert 1_{3}\right\rangle +\sqrt{\frac{1}{3}}e^{-i\pi
/3}\left\vert 1_{1}\right\rangle \left\vert \psi _{2,3}^{+}\right\rangle %
\right] .  \nonumber
\end{eqnarray}%
Then $Q_{1}$ is tuned back to its idle frequency of $5.367$ GHz and decoupled from $Q_{2}$
and $Q_{3}$, which remain at the working frequency and continue to interact
with each other. The state components $\left\vert 0_{2}\right\rangle
\left\vert 0_{3}\right\rangle $ and $\left\vert 1_{2}\right\rangle
\left\vert 1_{3}\right\rangle $ are eigenstates of the two-qubit interaction
Hamiltonian $H_{\text{e}}^{^{\prime }}=-\lambda \left(
S_{2}^{+}S_{3}^{-}+S_{2}^{-}S_{3}^{+}\right) $ with the zero eigenvalue,
while $\left\vert \psi _{2,3}^{+}\right\rangle $ is an eigenstate of $%
H_{\text{e}}^{^{\prime }}$ with the eigenvalue of $-\lambda $.
As a result, this
swapping interaction does not affect $\left\vert 0_{2}\right\rangle
\left\vert 0_{3}\right\rangle $ and $\left\vert 1_{2}\right\rangle
\left\vert 1_{3}\right\rangle $, but produces a phase shift $\lambda \tau
^{^{\prime }}$ to $\left\vert \psi _{2,3}^{+}\right\rangle $, with $\tau
^{^{\prime }}$ being the interaction time. With the choice $\tau ^{^{\prime
}}=\pi /3\lambda $, the two-qubit coupling $C_{2,3}$ cancels the phase factor $e^{-i\pi /3}$ associated
with $\left\vert \psi _{2,3}^{+}\right\rangle $, evolving the three qubits
to \cite{zou_pra2003}%
\begin{eqnarray}
&&\ \ \alpha \left[ \sqrt{\frac{2}{3}}e^{i\phi }\left\vert
1_{1}\right\rangle \left\vert 0_{2}\right\rangle \left\vert
0_{3}\right\rangle +\sqrt{\frac{1}{3}}\left\vert 0_{1}\right\rangle
\left\vert \psi _{2,3}^{+}\right\rangle \right] \\
&&\ \ +\beta \left[ \sqrt{\frac{2}{3}}\left\vert 0_{1}\right\rangle
\left\vert 1_{2}\right\rangle \left\vert 1_{3}\right\rangle +\sqrt{\frac{1}{3%
}}e^{i\phi }\left\vert 1_{1}\right\rangle \left\vert \psi
_{2,3}^{+}\right\rangle \right] ,  \nonumber
\end{eqnarray}%
where the phase $\phi $ is due to the frequency shift of $Q_{1}$ during
the $Q_{2}$-$Q_{3}$ interaction, which does not affect the reduced density
matrices for both $Q_{2}$ and $Q_{3}$, each of which in the basis $\left\{
\left\vert 0\right\rangle ,\left\vert 1\right\rangle \right\} $ is given by
\begin{equation}
\left(
\begin{array}{cc}
\frac{5}{6}\left\vert \alpha \right\vert ^{2}+\frac{1}{6}\left\vert \beta
\right\vert ^{2} & \frac{2}{3}\alpha \beta ^{\ast } \\
\frac{2}{3}\alpha ^{\ast }\beta & \frac{1}{6}\left\vert \alpha \right\vert
^{2}+\frac{5}{6}\left\vert \beta \right\vert ^{2}%
\end{array}%
\right) .
\end{equation}%
For the perfect UQCM, the fidelity of these two output copiers with respect
to the input state $\left\vert \psi _{\text{in}}\right\rangle $ is $5/6$,
irrespective of the probability amplitudes $\alpha $ and $\beta $ associated
with the components $\left\vert 0\right\rangle $ and $\left\vert
1\right\rangle $. 
Due to the nonuniform qubit-resonator couplings and the existence of the direct but also nonuniform qubit-qubit couplings in our device \cite{zhong_prl2016,song_prl2017,xu_prl2018,ning_prl2019,song_science2019}, each qubit is asymmetrically coupled to the other two qubits with the effective coupling strengths slightly different from $\lambda$. In order to produce the optimal outcome, the coupling operations $C_{123}$ and $C_{23}$ are calibrated simultaneously, and consequently, the optimal coupling times $\tau$ and $\tau^\prime$ that deviate from the values for the ideal case are 40.8 ns and 69.5 ns, respectively.
After the copy process, $Q_2$ and $Q_3$ are tuned back to their idle frequencies of $5.223$ GHz and $5.311$ GHz, respectively.

\subsection{Characterization of performance.}
We characterize the performance of the UQCM by preparing different input
states $\{\left\vert 0_{1}\right\rangle$, $\left( \left\vert
0_{1}\right\rangle +i\left\vert 1_{1}\right\rangle \right) /\sqrt{2}$, $\left(\left\vert
0_{1}\right\rangle-i\left\vert 1_{1}\right\rangle \right)/\sqrt{2}$, $\left(\left\vert
0_{1}\right\rangle+\left\vert 1_{1}\right\rangle \right)/\sqrt{2}$, $\left(\left\vert
0_{1}\right\rangle-\left\vert 1_{1}\right\rangle \right)/\sqrt{2}$,
$\left\vert 1_{1}\right\rangle \} $, and measuring the corresponding
output states of $Q_{2}$ and $Q_{3}$ through quantum state tomography (See Supplementary Note 7).
Note that, for the tomography in our experiment, the readout calibration is performed to correct the measured probabilities for 
the qubit states according to the qubits' $\left\vert 0\right\rangle$- and $\left\vert 1\right\rangle$-state 
measurement fidelities (see Supplementary Table 1 and Note 6).
The measured density matrices for the clones of the
above-mentioned six input states are respectively displayed in Fig. \ref{f2}a-f, where the upper and lower panels denote the measured output
density matrices of $Q_{2}$ and $Q_{3}$, respectively. The fidelities of the
output states of $Q_{2}$ ($Q_{3}$) to these six ideal input states, defined
as $F=\left\langle \psi _{\text{in}}\right\vert \rho _{\text{out}}\left\vert
\psi _{\text{in}}\right\rangle $, are respectively $0.784\pm0.002$ ($0.824\pm0.002$), $0.786\pm0.001$ ($0.818\pm0.003$), $0.784\pm0.002$ ($0.832\pm0.002$), $0.788\pm0.002$ ($0.831\pm0.001$), $0.786\pm0.002$ ($0.832\pm0.001$), and $0.785\pm0.002$ ($0.832\pm0.001$), where $\rho _{\text{out}}$ denotes the measured density matrix for the
corresponding output clone. Each of these fidelities is close to the optimal
value 5/6, confirming the performance of the UQCM is independent of the
input state. The slight difference between the output states of the two copy
qubits is mainly due to direct qubit-qubit couplings. These nonuniform
couplings also make the qualities of the output states slightly depend on
the input state. We note that for each of the six input states, the output state of $Q_3$ has a fidelity very close to the theoretical upper bound. This is partly due to the asymmetry between the two clones. The other reason is that the qubit-qubit couplings during the copy process partly protect the qubits from dephasing, so that the real $T_2$ times of the qubits coupled at the working frequency are longer than the corresponding results listed in Supplementary Table 1, which are measured without qubit-qubit couplings \cite{song_prl2017,xu_prl2018}.

\begin{figure}[t]
	\centering
	\includegraphics[width=3.2in]{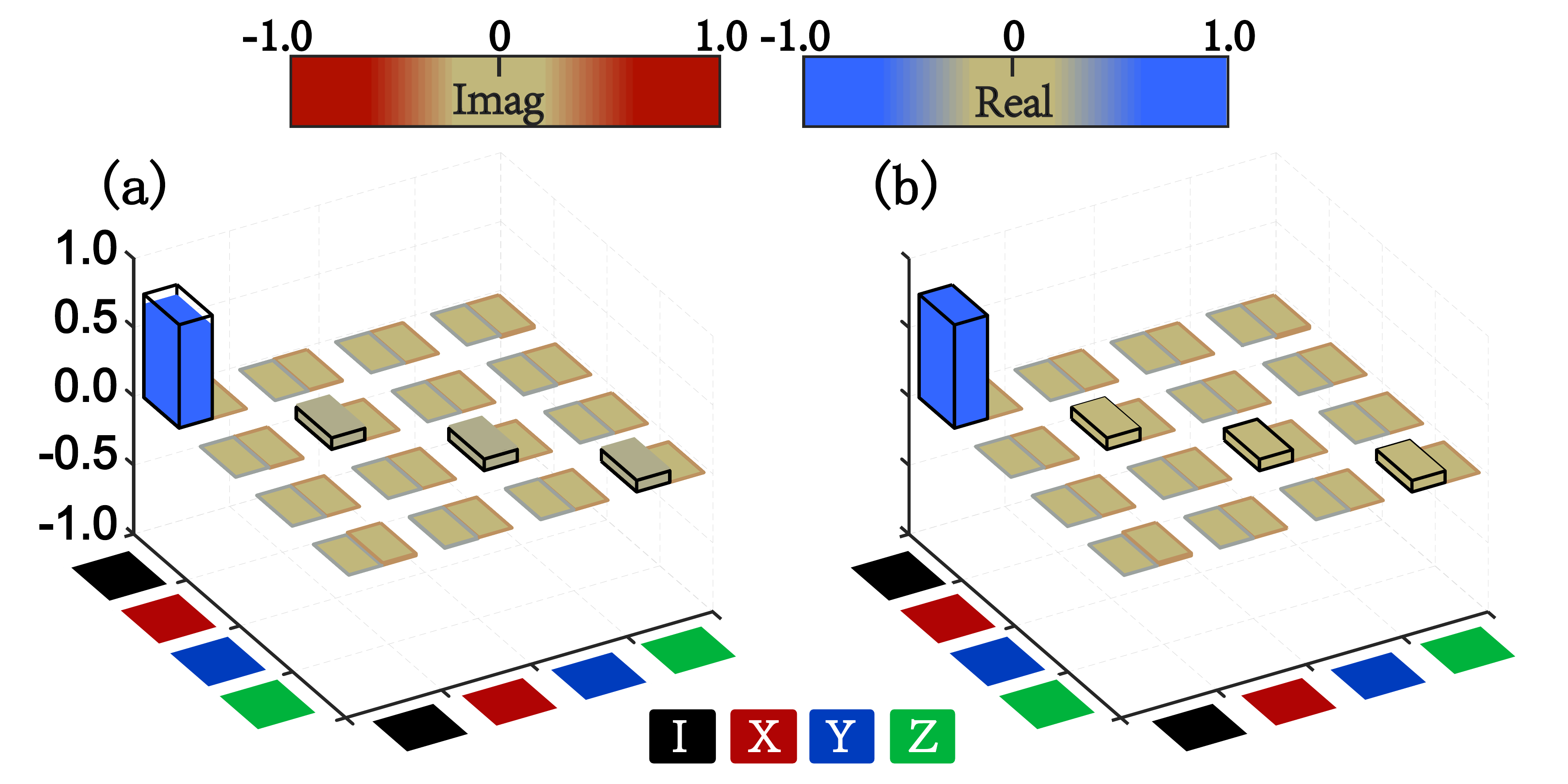}
	\caption{Process tomography. \textbf{a} Measured process matrix
		associated with the output state of $Q_{2}$. \textbf{b} Measured process matrix
		associated with $Q_{3}$'s output state. The black wire frames denote the
		corresponding process matrix elements of the perfect UQCM.}
	\label{f3}
\end{figure}

\begin{figure*}[t]
	\centering
	\includegraphics[width=7in]{Figure4_concurrence_noreverse_V1-eps-converted-to.pdf}
	\caption{Measured $Q_{1}$-$Q_{2}$ and $Q_{1}$-$Q_{3}$ output
		density matrices for input states: \textbf{a} $\left\vert 0_{1}\right\rangle $; \textbf{b} $%
		\left( \left\vert 0_{1}\right\rangle +i\left\vert 1_{1}\right\rangle \right)
		/\sqrt{2}$; \textbf{c} $\left( \left\vert 0_{1}\right\rangle -i\left\vert 1_{1}\right\rangle \right)
		/\sqrt{2}$; \textbf{d} $\left( \left\vert 0_{1}\right\rangle +\left\vert 1_{1}\right\rangle \right)
		/\sqrt{2}$; \textbf{e} $\left( \left\vert 0_{1}\right\rangle -\left\vert
		1_{1}\right\rangle \right) /\sqrt{2}$; \textbf{f} $\left\vert 1_{1}\right\rangle $.
		For clarity, a single-qubit z-axis rotation is numerically applied to cancel
		the phase of the extra phase accumulated due to the qubits' frequency
		shift. The measured $Q_{1}$-$Q_{2}$ and $Q_{1}$-$Q_{3}$ output density
		matrices are displayed in the upper and lower panels, respectively. The black wire frames denote the
		corresponding density matrices produced by the perfect UQCM.}
	\label{f4}
\end{figure*}
To further examine the performance of the UQCM, we perform the quantum
process tomography (See Supplementary Note 7), achieved by preparing the above mentioned six distinct
input states, and measuring them and the corresponding output states of $%
Q_{2}$ and $Q_{3}$ through quantum state tomography. The measured process
matrices associated with the output states of $Q_{2}$ and $Q_{3}$, $\chi _{%
\text{meas,2}}$ and $\chi _{\text{meas,3}}$, are respectively presented in
Fig. \ref{f3}a and \ref{f3}b, respectively. The fidelities of $\chi _{\text{meas,2}}$ and $%
\chi _{\text{meas,3}}$ with respect to the ideal cloning process $\chi _{%
\text{id}}$, defined as $F=Tr\left( \chi _{\text{meas}}\chi _{\text{id}%
}\right) $, are $0.679\pm0.001$ and $0.743\pm0.002$, respectively. These process fidelities are close to
the result of the perfect UQCM, 0.75, demonstrating a good quantum control
over the multiqubit-resonator system.

\subsection{Demonstration of universal entanglement behavior}
The nonclassical correlations between the original input qubit and the clones play an essential role in implementation of the UQCM and represents one of the most fundamental differences between universal quantum and classical cloning, but have not been quantitatively investigated. Characterization of these correlations is important for understanding the quantum behavior of the UQCM. We find that the degree of the entanglement between each output clone and the original input qubit, quantified by concurrence \cite{wootters_prl1998}, is 2/3 for an ideal UQCM, which is independent of the input state (see Supplementary Note 1).
To detect these nonclassical correlations, we respectively
measure the joint $Q_{1}$-$Q_{2}$ and $Q_{1}$-$Q_{3}$ output density
matrices. 
The results for
the six input states $\{\left\vert 0_{1}\right\rangle$, $\left( \left\vert
0_{1}\right\rangle +i\left\vert 1_{1}\right\rangle \right) /\sqrt{2}$, $\left(\left\vert
0_{1}\right\rangle-i\left\vert 1_{1}\right\rangle \right)/\sqrt{2}$, $\left(\left\vert
0_{1}\right\rangle+\left\vert 1_{1}\right\rangle \right)/\sqrt{2}$, $\left(\left\vert
0_{1}\right\rangle-\left\vert 1_{1}\right\rangle \right)/\sqrt{2}$,
$\left\vert 1_{1}\right\rangle \} $ are displayed in
Fig. \ref{f4}a-f, where the upper and lower panels correspond to the joint $Q_{1}$-$%
Q_{2}$ and $Q_{1}$-$Q_{3}$ output density matrices, respectively. For
clarity of display, a single-qubit z-axis rotation is numerically applied to
cancel the extra phase produced by the qubits' frequency shift, which
does not affect the entanglement. The output $Q_{1}$-$Q_{2}$ concurrences
associated with these six input states are $0.667\pm0.008$, $0.557\pm0.005$, $0.554\pm0.006$, $0.556\pm0.007$, $0.569\pm0.007$, and $0.590\pm0.008$, while the
corresponding output $Q_{1}$-$Q_{3}$ concurrences are $0.565\pm0.011$, $0.453\pm0.008$, $0.457\pm0.005$, $0.468\pm0.005$, $0.459\pm0.007$, and $0.511\pm0.006$,
respectively. 
As a nonzero concurrence indicates the existence of entanglement, these results unambiguously demonstrate that each of the two copy qubits is highly entangled with the original qubit, and the two copy qubits are also entangled.
When the original qubit is traced out, the concurrence between the two copy qubits is 1/3 for an ideal UQCM (see Supplementary Note 1), which is also input-state independent \cite{buzek_pra1997}.
To verify this entanglement, we perform the joint $Q_{2}$-$Q_{3}$ output
state tomography. 
The reconstructed joint $Q_{2}$-$Q_{3}$ density matrices for the six input
states are displayed in Supplementary Note 5. The $Q_{2}$-$Q_{3}$
concurrences associated with these six measured density matrices are
$0.236\pm0.012$, $0.076\pm0.007$, $0.076\pm0.006$, $0.079\pm0.007$, $0.088\pm0.005$, and $0.180\pm0.010$, respectively. These results indicate that for the input
superposition state the output $Q_{2}$-$Q_{3}$ entanglement is much more
affected by the decoherence effect compared to the case with input $%
\left\vert 0_{1}\right\rangle $- or $\left\vert 1_{1}\right\rangle $-state.
This can be interpreted as follows. Since the direct couplings between the qubits are asymmetric, after the copy process, different state components of the final three-qubit state will accumulate different phases. For each input superposition state, the $Q_2-Q_3$ concurrence on these phases is a cosine function of these phases, which is very different from the modulation of the fidelity of the output state of each copy qubit (see Supplementary Note 2). With the present system parameters, the value of the modulation function of the $Q_2-Q_3$ concurrence is much smaller than that of the fidelity of the output state of each copy qubit, which approximates to the maximum. Consequently, the output $Q_2-Q_3$ entanglement for the input superposition state is much smaller than that for the input $\left\vert0\right\rangle$- or $\left\vert1\right\rangle$-state; while the output state fidelity of each qubit is almost input-state-independent. The existence of concurrence between any
two of the three qubits confirm they are in a genuine three-particle entangled
state, revealing the fundamental difference between a quantum cloning
process and a classical one.

\section{DISCUSSION}
We have demonstrated universal cloning of an arbitrary state of an individual qubit with a circuit QED setup, where all the quantum operations necessary for constructing a UQCM network are deterministically realized. We characterize the performance of the UQCM by quantum state tomography, confirming the universality of the copying process. We measure the entanglement between each copy qubit and the original qubit, with the results being in well agreement with the theoretical prediction that this entanglement is input-state-independent and represents a universal quantum behavior of the UQCM. We further measure the entanglement between the two clones, verifying the existence of true three-particle entanglement at the output. These results underline the fact that the universal entanglement behavior underlies the performance of the UQCM.

\section{DATA AVAILABILITY}
All data needed to evaluate the conclusions in the
paper are present in the paper and/or the Supplementary Materials.
Additional data related to this paper may be requested from the authors.

\section{CODE AVAILABILITY}
All codes used in the paper are available from the corresponding authors upon reasonable request.

\section{ACKNOWLEDGMENTS}
We thank Haohua Wang at Zhejiang University for technical support. This work was supported by the National Natural Science
Foundation of China (Grant No. 11674060, No. 11874114, No. 11875108, No. 11934018, No. 11904393 and No. 92065114),
the Strategic Priority Research Program of Chinese Academy of Sciences
(Grant No. XDB28000000), and the Natural Science Foundation of Fujian Province under Grant No. 2018J01412.

\section{AUTHOR CONTRIBUTIONS}
S.-B.Z. conceived the experiment. Z.-B.Y., P.-R.H., X.-J.H., and K.X. performed the experiment and analyzed the data with the assistance of W. N.
H.L. and D.Z. provided the devices used for the experiment. S.-B.Z., Z.-B.Y., K.X. and H.F. wrote the manuscript with feedbacks from all authors.

\section{COMPETING INTERESTS}
The authors declare that they have no competing interests.

\end{document}


\title{Supplementary Material for: Experimental demonstration of entanglement-enabled universal quantum cloning in a circuit}

\author{Zhen-Biao Yang$^{1}$}
\author{Pei-Rong Han$^{1}$}
\author{Xin-Jie Huang$^{1}$}
\author{Wen Ning$^{1}$}
\author{Hekang Li$^{2}$}
\author{Kai Xu$^{2,3}$}
\email{kaixu@iphy.ac.cn}
\author{Dongning Zheng$^{2,3}$}
\author{Heng Fan$^{2,3}$}
\email{hfan@iphy.ac.cn}
\author{Shi-Biao Zheng$^{1}$}
\email{t96034@fzu.edu.cn}
\affiliation{1.Fujian Key Laboratory of Quantum Information and Quantum Optics, College of Physics and Information Engineering, Fuzhou University, Fuzhou, Fujian 350108, China}
\affiliation{2.Institute of Physics and Beijing National Laboratory for Condensed Matter Physics, Chinese Academy of Sciences, Beijing 100190, China}
\affiliation{3.CAS Center for Excellence in Topological Quantum Computation, University of Chinese Academy of Sciences, Beijing 100190, China}

\maketitle

\begin{center}
	\bf{SUPPLEMENTARY NOTE 1: Quantum entanglement in the output state of an ideal UQCM}
\end{center}


When one of the copying qubits is traced out, the joint density matrix for the
other copy qubit and the original qubit at the output of the UQCM, in the basis $\left\{ \left\vert
0_{1}0_{k}\right\rangle ,\left\vert 0_{1}1_{k}\right\rangle ,\left\vert
1_{1}0_{k}\right\rangle ,\left\vert 1_{1}1_{k}\right\rangle \right\}$ ($k=2$ or $3$),
is given by

\begin{small}
\begin{equation}
\rho =\left(
\begin{array}{cccc}
\frac{1}{6}\left\vert \alpha \right\vert ^{2} & \frac{1}{3}\alpha \beta
^{\ast } & \frac{1}{6}\alpha \beta ^{\ast } & 0 \\
\frac{1}{3}\alpha ^{\ast }\beta & \frac{1}{6}\left\vert \alpha \right\vert
^{2}+\frac{2}{3}\left\vert \beta \right\vert ^{2} & \frac{1}{3} & \frac{1}{6}%
\alpha \beta ^{\ast } \\
\frac{1}{6}\alpha ^{\ast }\beta & \frac{1}{3} & \frac{2}{3}\left\vert \alpha
\right\vert ^{2}+\frac{1}{6}\left\vert \beta \right\vert ^{2} & \frac{1}{3}%
\alpha \beta ^{\ast } \\
0 & \frac{1}{6}\alpha ^{\ast }\beta & \frac{1}{3}\alpha ^{\ast }\beta &
\frac{1}{6}\left\vert \beta \right\vert ^{2}%
\end{array}%
\right) .\text{ }
\end{equation}%
\end{small}
The corresponding $\stackrel{\sim }{\rho }$ matrix, defined as $\stackrel{\sim }{\rho }=\rho \left( \sigma
_{y}\otimes \sigma _{y}\right) \rho ^{\ast }\left( \sigma _{y}\otimes \sigma
_{y}\right)$, is
\begin{widetext}
\begin{eqnarray}
\stackrel{\sim }{\rho }=\left(
\begin{array}{cccc}
-\frac{1}{9}\left\vert \alpha \beta \right\vert ^{2} & \frac{1}{9}\alpha
\beta ^{\ast }\left( 2\left\vert \alpha \right\vert ^{2}+\left\vert \beta
\right\vert ^{2}\right) & \frac{1}{9}\alpha \beta ^{\ast }\left( \left\vert
\alpha \right\vert ^{2}+2\left\vert \beta \right\vert ^{2}\right) & \frac{-1%
}{9}\left( \alpha \beta ^{\ast }\right) ^{2} \\
-\frac{1}{9}\alpha ^{\ast }\beta \left( \left\vert \alpha \right\vert
^{2}+2\left\vert \beta \right\vert ^{2}\right) & \frac{2}{9}\left\vert
\alpha \right\vert ^{4}+\frac{5}{9}\left\vert \alpha \beta \right\vert ^{2}+%
\frac{2}{9}\left\vert \beta \right\vert ^{4} & \frac{1}{9}\left\vert \alpha
\right\vert ^{4}+\frac{4}{9}\left\vert \alpha \beta \right\vert ^{2}+\frac{4%
}{9}\left\vert \beta \right\vert ^{4} & \frac{-1}{9}\alpha \beta ^{\ast
}\left( \left\vert \alpha \right\vert ^{2}+2\left\vert \beta \right\vert
^{2}\right) \\
\frac{-1}{9}\alpha ^{\ast }\beta \left( 2\left\vert \alpha \right\vert
^{2}+\left\vert \beta \right\vert ^{2}\right) & \frac{4}{9}\left\vert \alpha
\right\vert ^{4}+\frac{4}{9}\left\vert \alpha \beta \right\vert ^{2}+\frac{1%
}{9}\left\vert \beta \right\vert ^{4} & \frac{2}{9}\left\vert \alpha
\right\vert ^{4}+\frac{5}{9}\left\vert \alpha \beta \right\vert ^{2}+\frac{2%
}{9}\left\vert \beta \right\vert ^{4} & \frac{-1}{9}\alpha \beta ^{\ast
}\left( 2\left\vert \alpha \right\vert ^{2}+\left\vert \beta \right\vert
^{2}\right) \\
\frac{-1}{9}\left( \alpha ^{\ast }\beta \right) ^{2} & \frac{1}{9}\alpha
^{\ast }\beta \left( 2\left\vert \alpha \right\vert ^{2}+\left\vert \beta
\right\vert ^{2}\right) & \frac{1}{9}\alpha ^{\ast }\beta \left( \left\vert
\alpha \right\vert ^{2}+2\left\vert \beta \right\vert ^{2}\right) & \frac{-1%
}{9}\left\vert \alpha \beta \right\vert ^{2}%
\end{array}%
\right).
\end{eqnarray}
\end{widetext}
The four eigenvalues of $\stackrel{\sim }{\rho }$ in the decreasing order
are $\lambda _{1}=4/9$ and $\lambda _{2}=\lambda _{3}=\lambda
_{4}=0$, respectively. The corresponding concurrence \cite{wootters_prl1998}, defined as $C=\max \{\sqrt{%
\lambda _{1}}-\sqrt{\lambda _{2}}-\sqrt{\lambda _{3}}-\sqrt{\lambda _{4}}$, $%
0\}$, is 2/3, which is independent of the input state.

After tracing over the original qubit, the joint density matrix of $Q_{2}$
and $Q_{3}$ at the output of the UQCM, in the basis $\left\{ \left\vert
0_{2}0_{3}\right\rangle ,\left\vert 0_{2}1_{3}\right\rangle ,\left\vert
1_{2}0_{3}\right\rangle ,\left\vert 1_{2}1_{3}\right\rangle \right\} $,
reads as

\begin{equation}
\rho=\left(
\begin{array}{cccc}
\frac{2}{3}\left\vert \alpha \right\vert ^{2} & \frac{1}{3}\alpha \beta
^{\ast } & \frac{1}{3}\alpha \beta ^{\ast } & 0 \\
\frac{1}{3}\alpha ^{\ast }\beta & \frac{1}{6} & \frac{1}{6} & \frac{1}{3}%
\alpha \beta ^{\ast } \\
\frac{1}{3}\alpha ^{\ast }\beta & \frac{1}{6} & \frac{1}{6} & \frac{1}{3}%
\alpha \beta ^{\ast } \\
0 & \frac{1}{3}\alpha ^{\ast }\beta & \frac{1}{3}\alpha ^{\ast }\beta &
\frac{2}{3}\left\vert \beta \right\vert ^{2}%
\end{array}%
\right) .\text{ }
\end{equation}%
The matrix $\stackrel{\sim }{\rho }$ is
\begin{widetext}
\begin{eqnarray}
\stackrel{\sim }{\rho }=\left(
\begin{array}{cccc}
\frac{2}{9}\left\vert \alpha \beta \right\vert ^{2} & \frac{1}{9}\alpha
\beta ^{\ast }\left( \left\vert \beta \right\vert ^{2}-\left\vert \alpha
\right\vert ^{2}\right) & \frac{1}{9}\alpha \beta ^{\ast }\left( \left\vert
\beta \right\vert ^{2}-\left\vert \alpha \right\vert ^{2}\right) & \frac{-2}{%
9}\left( \alpha \beta ^{\ast }\right) ^{2} \\
\frac{1}{9}\alpha ^{\ast }\beta \left( \left\vert \beta \right\vert
^{2}-\left\vert \alpha \right\vert ^{2}\right) & \frac{1}{18}-\frac{2}{9}%
\left\vert \alpha \beta \right\vert ^{2} & \frac{1}{18}-\frac{2}{9}%
\left\vert \alpha \beta \right\vert ^{2} & \frac{1}{9}\alpha \beta ^{\ast
}\left( \left\vert \alpha \right\vert ^{2}-\left\vert \beta \right\vert
^{2}\right) \\
\frac{1}{9}\alpha ^{\ast }\beta \left( \left\vert \beta \right\vert
^{2}-\left\vert \alpha \right\vert ^{2}\right) & \frac{1}{18}-\frac{2}{9}%
\left\vert \alpha \beta \right\vert ^{2} & \frac{1}{18}-\frac{2}{9}%
\left\vert \alpha \beta \right\vert ^{2} & \frac{1}{9}\alpha \beta ^{\ast
}\left( \left\vert \alpha \right\vert ^{2}-\left\vert \beta \right\vert
^{2}\right) \\
\frac{-2}{9}\left( \alpha ^{\ast }\beta \right) ^{2} & \frac{1}{9}\alpha
^{\ast }\beta \left( \left\vert \alpha \right\vert ^{2}-\left\vert \beta
\right\vert ^{2}\right) & \frac{1}{9}\alpha ^{\ast }\beta \left( \left\vert
\alpha \right\vert ^{2}-\left\vert \beta \right\vert ^{2}\right) & \frac{2}{9%
}\left\vert \alpha \beta \right\vert ^{2}%
\end{array}%
\right),
\end{eqnarray}
\end{widetext}
the four eigenvalues of which in the decreasing order
are $\lambda _{1}=1/9$ and $\lambda _{2}=\lambda _{3}=\lambda
_{4}=0$, respectively, corresponding to a concurrence of 1/3, which is also
input-state-independent.

\begin{table*}[!htb]
	\centering
	\begin{tabular}{cccccccccc}
		\hline
		\hline
		&$\omega_{j}/2\pi$ (GHz)& $g_j/2\pi$ (MHz) &$\lambda_{j,j+1}/2\pi$ (MHz)&$T_{1,j}$ ($\mu$s)&$T_{2,j}^*$ ($\mu$s)&$T_{2,j}^{\textrm{SE}}$ ($\mu$s)&$1/\kappa^r_j (ns)$&$F_{0,j}$&$F_{1,j}$\\
		\hline
		$Q_1$&5.367&20.0&0.069&24.0(15.4)&2.2(2.4)&6.5(6.9)&315&0.984&0.925\\
		$Q_2$&5.223&20.8&0.553&25.8(28.5)&0.7(0.9)&6.2(7.1)&219&0.988&0.939\\
		$Q_3$&5.311&19.9&-0.066&22.7(14.4)&2.6(2.9)&8.0(8.9)&203&0.986&0.921\\
		\hline
		\hline
	\end{tabular}
	\caption{\label{table1} \textbf{Qubits characteristics.} $\omega_{j}/2\pi$ is the idle frequency of $Q_j$ where single-qubit rotations and qubit state tomography are performed. $g_j$ is the coupling strength between $Q_j$ and the bus resonator. $\lambda_{j,j+1}$ is the magnitude of crosstalk qubit-qubit coupling between $Q_j$ and $Q_{j+1}$ beyond that induced by the resonator. $T_{1,j}$, $T_{2,j}^*$, and $T_{2,j}^{\textrm{SE}}$ are the energy relaxation time, the Ramsey dephasing time (Gaussian decay), and the dephasing time (Gaussian decay) with spin-echo, respectively, all of them (those in parentheses) are measured at the idle (working) point of $Q_j$.
		$\kappa_{j}^{r}$ represents the leakage rate of $Q_j$'s readout resonator $R_j$. $F_{0,j}$ ($F_{1,j}$) is the probability of correctly detecting $Q_j$ in $\vert 0\rangle$ ($\vert 1\rangle$) when it is prepared in $\vert 0\rangle$ ($\vert 1\rangle$).}
\end{table*}
\begin{center}
	\bf{SUPPLEMENTARY NOTE 2: Effects of asymmetric direct couplings}
\end{center}


With direct asymmetric couplings between the qubits being considered, the
system finally evolves to the state
\begin{widetext}
\begin{eqnarray}
&&\alpha \left[ \sqrt{\frac{2}{3}}k_{1}e^{i\theta }\left\vert
1_{1}0_{2}0_{3}\right\rangle +\sqrt{\frac{1}{6}}\left\vert
0_{1}\right\rangle (k_{2}e^{i\phi _{2}}\left\vert 1_{2}0_{3}\right\rangle
+k_{3}e^{i\phi _{3}}\left\vert 0_{2}1_{3}\right\rangle )\right]  \\
&&+\beta \left[ \sqrt{\frac{2}{3}}k_{4}e^{i\phi _{4}}\left\vert
0_{1}1_{2}1_{3}\right\rangle +\sqrt{\frac{1}{6}}e^{i\theta }\left\vert
1_{1}\right\rangle (k_{5}e^{i\phi _{5}}\left\vert 1_{2}0_{3}\right\rangle
+k_{6}e^{i\phi _{6}}\left\vert 0_{2}1_{3}\right\rangle )\right] .  \nonumber
\end{eqnarray}%
\end{widetext}
The reduced density operator for $Q_{2}$ and $Q_{3}$ after tracing out $Q_{1}
$ is given by%
\begin{equation}
\ \left\vert \psi _{1}\right\rangle \left\langle \psi _{1}\right\vert
+\left\vert \psi _{2}\right\rangle \left\langle \psi _{2}\right\vert ,
\end{equation}%
where%
\begin{eqnarray}
&&\ \ \left\vert \psi _{1}\right\rangle =\sqrt{\frac{2}{3}}\alpha
k_{1}e^{i\theta }\left\vert 0_{2}0_{3}\right\rangle \\ \nonumber
&&\ \ +\sqrt{\frac{1}{6}}\beta
e^{i\theta }(k_{5}e^{i\phi _{5}}\left\vert 1_{2}0_{3}\right\rangle
+k_{6}e^{i\phi _{6}}\left\vert 0_{2}1_{3}\right\rangle ),
\end{eqnarray}%
and%
\begin{small}
\begin{equation}
\ \left\vert \psi _{2}\right\rangle =\sqrt{\frac{1}{6}}\alpha (k_{2}e^{i\phi
	_{2}}\left\vert 1_{2}0_{3}\right\rangle +k_{3}e^{i\phi _{3}}\left\vert
0_{2}1_{3}\right\rangle )+\sqrt{\frac{2}{3}}\beta k_{4}e^{i\phi
	_{4}}\left\vert 1_{2}1_{3}\right\rangle .
\end{equation}%
\end{small}
The concurrence between $Q_{2}$ and $Q_{3}$ is given by%
\begin{equation}
{\cal E}_{2,3}=\frac{1}{3}\left\vert \left\vert \alpha \right\vert
^{2}k_{2}k_{3}e^{i\left( \phi _{2}-\phi _{3}\right) }+\left\vert \beta
\right\vert ^{2}k_{5}k_{6}e^{i\left( \phi _{5}-\phi _{6}\right) }\right\vert
.
\end{equation}%
For $k_{j}\approx 1$ with $j=1$ to $5$, we have
\begin{equation}
{\cal E}_{2,3}\approx \frac{1}{3}\left\vert \left\vert \alpha \right\vert
^{2}e^{i\left( \phi _{2}-\phi _{3}\right) }+\left\vert \beta \right\vert
^{2}e^{i\left( \phi _{5}-\phi _{6}\right) }\right\vert .
\end{equation}%
When $\beta =0$ or $1$, we have ${\cal E}_{2,3}\approx \frac{1}{3}$. For $%
\left\vert \beta \right\vert =1/\sqrt{2}$, ${\cal E}_{2,3}\approx \frac{1}{3}%
\cos \delta $, where $\delta =\left( \phi _{2}+\phi _{6}-\phi _{3}-\phi
_{3}\right) /2$. This implies that the output $Q_{2}$-$Q_{3}$ entanglement
for the input superposition state may be much smaller than that for the
input $\left\vert 0\right\rangle $- or $\left\vert 1\right\rangle $-state.

When $Q_{1}$ and $Q_{2}$ are traced out, the reduced density matrix for $%
Q_{3}$ is given by%
\begin{widetext}
\begin{equation}
\rho _{3}=\left(
\begin{array}{cc}
\frac{2}{3}\left\vert \alpha k_{1}\right\vert ^{2}+\frac{1}{6}\left\vert
\alpha k_{2}\right\vert ^{2}+\frac{1}{6}\left\vert \beta k_{5}\right\vert
^{2} & \frac{1}{3}\alpha \beta ^{\ast }k_{1}k_{6}e^{-i\phi _{6}}+\frac{1}{3}%
\alpha \beta ^{\ast }k_{2}k_{4}e^{i\left( \phi _{2}-\phi _{4}\right) } \\
\frac{1}{3}\alpha ^{\ast }\beta k_{1}k_{6}e^{i\phi _{6}}+\frac{1}{3}\alpha
^{\ast }\beta k_{2}k_{4}e^{-i\left( \phi _{2}-\phi _{4}\right) } & \frac{1}{6%
}\left\vert \alpha k_{3}\right\vert ^{2}+\frac{1}{6}\left\vert \beta
k_{6}\right\vert ^{2}+\frac{2}{3}\left\vert \beta k_{4}\right\vert ^{2}%
\end{array}%
\right) .
\end{equation}%
\end{widetext}
For $k_{j}\approx 1$, we have%
\begin{equation}
\rho _{3}\approx \left(
\begin{array}{cc}
\frac{5}{6}\left\vert \alpha \right\vert ^{2}+\frac{1}{6}\left\vert \beta
\right\vert ^{2} & \frac{1}{3}\alpha \beta ^{\ast }\left[ e^{-i\phi
	_{6}}+e^{i\left( \phi _{2}-\phi _{4}\right) }\right]  \\
\frac{1}{3}\alpha ^{\ast }\beta \left[ e^{i\phi _{6}}+e^{-i\left( \phi
	_{2}-\phi _{4}\right) }\right]  & \frac{1}{6}\left\vert \alpha \right\vert
^{2}+\frac{5}{6}\left\vert \beta \right\vert ^{2}%
\end{array}%
\right) .
\end{equation}%
The corresponding fidelity is%
\begin{widetext}
\begin{eqnarray}
F &\approx &\left(
\begin{array}[t]{cc}
\alpha ^{\ast } & \beta ^{\ast }%
\end{array}%
\right) \left(
\begin{array}{cc}
\frac{5}{6}\left\vert \alpha \right\vert ^{2}+\frac{1}{6}\left\vert \beta
\right\vert ^{2} & \frac{1}{3}\alpha \beta ^{\ast }\left[ e^{-i\phi
	_{6}}+e^{i\left( \phi _{2}-\phi _{4}\right) }\right]  \\
\frac{1}{3}\alpha ^{\ast }\beta \left[ e^{i\phi _{6}}+e^{-i\left( \phi
	_{2}-\phi _{4}\right) }\right]  & \frac{1}{6}\left\vert \alpha \right\vert
^{2}+\frac{5}{6}\left\vert \beta \right\vert ^{2}%
\end{array}%
\right) \left(
\begin{array}{c}
\alpha  \\
\beta
\end{array}%
\right)  \\
&=&\left\vert \alpha \right\vert ^{2}\left( \frac{5}{6}\left\vert \alpha
\right\vert ^{2}+\frac{1}{6}\left\vert \beta \right\vert ^{2}\right) +\frac{2%
}{3}\left\vert \alpha \beta \right\vert ^{2}\left[ \cos \phi _{6}+\cos
\left( \phi _{2}-\phi _{4}\right) \right] +\left\vert \beta \right\vert
^{2}\left( \frac{1}{6}\left\vert \alpha \right\vert ^{2}+\frac{5}{6}%
\left\vert \beta \right\vert ^{2}\right) .  \nonumber
\end{eqnarray}%
\end{widetext}
When $\beta =0$ or $1$, we have $F\approx \frac{5}{6}$. For $\left\vert
\beta \right\vert =1/\sqrt{2}$, $F\approx \frac{1}{2}+\frac{1}{6}\left[ \cos
\phi _{6}+\cos \left( \phi _{2}-\phi _{4}\right) \right] $. When $\phi
_{6}\approx 2m\pi $ and $\phi _{2}-\phi _{4}\approx 2n\pi $, with $m$ and $n$
being integers, the fidelities for the input superposition states are almost
the same as those for the input $\left\vert 0\right\rangle $- or $\left\vert
1\right\rangle $-state.

\begin{figure*}[t]
	\centering
	
	\includegraphics[width=0.61\textwidth,clip=True]{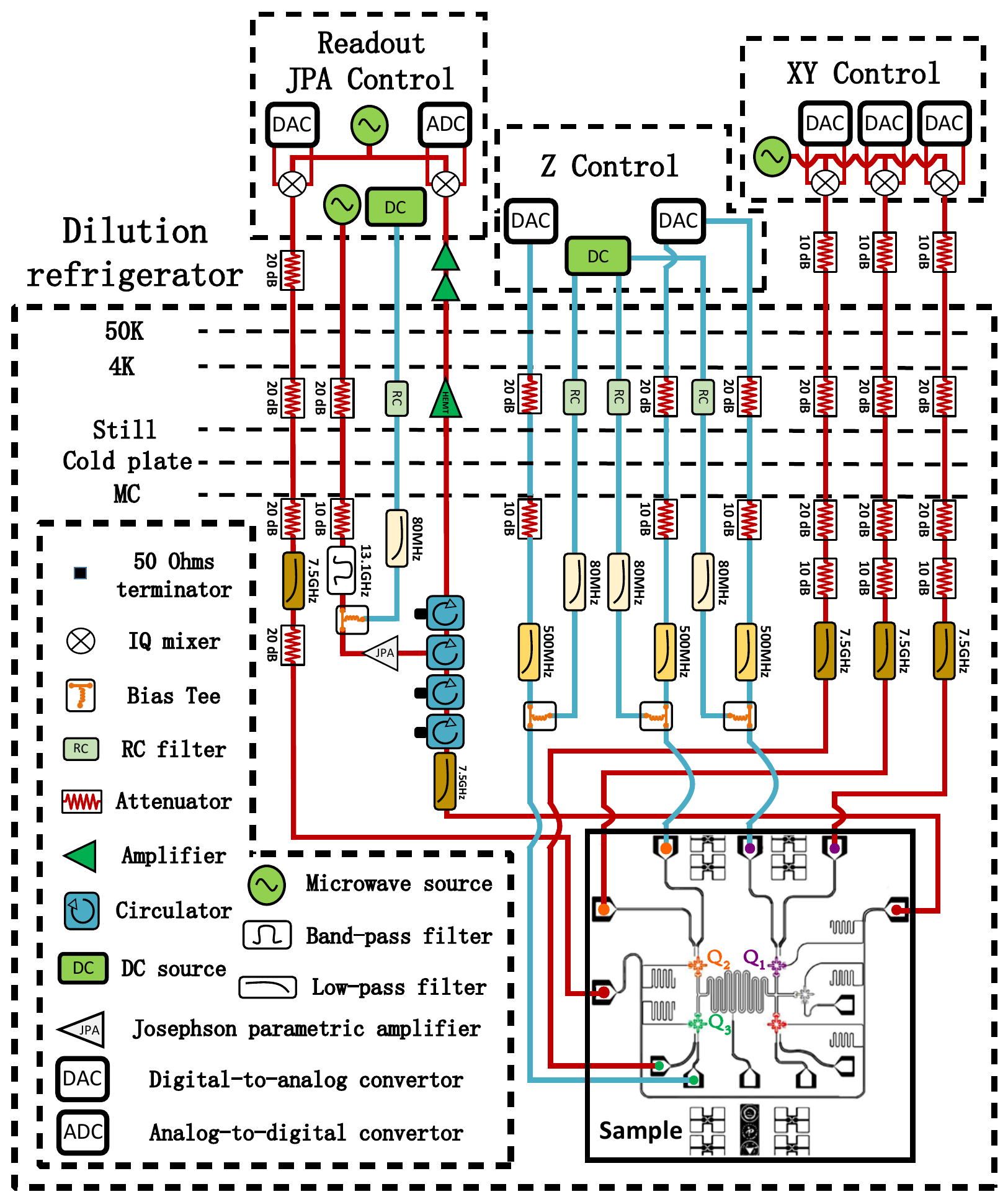}
	\caption{\label{cloning_wiring} \footnotesize{Sketch of the device and experimental setup. At the right-bottom side: the device (sample) consists of five
			frequency-tunable superconducting Xmon qubits, denoted by $Q_{j}$ with $j=1$ to $%
			5 $, coupled to a bus resonator, labeled as $R$. $Q_{1}$ acts as the input
			qubit, whose state is to be copied onto $Q_{2}$ and $Q_{3}$. $Q_{4}$ and $%
			Q_{5}$ are unused and not marked here. All the electronics and wiring is shown, from right to left, for the qubit XY control, the qubit Z control, the qubit readout and JPA control, respectively. Each qubit XY control is produced by the mixing of the signals of two independent DAC channels I/Q and a microwave source, and used for the fast qubit flipping.  Each qubit Z control has two channels of signals: the Z pulse control directly produced by a DAC channel for the fast frequency tuning of the qubit and the DC control provided by a DC source for resetting its operating point. The qubit readout signal is provided also by the sideband mixing of the signals of two independent DAC channels I/Q and a microwave source, outputting a three-tone microwave pulse that targets the readout resonators of all the qubits. The output signal is amplified sequentially by JPA, HEMT and room temperature amplifiers, before being captured and demodulated by the ADC. The JPA control is provided by the pumping of an independent microwave signal with its amplification band being tunable with a DC bias applied to it. Each control line is fed with well-designed attenuators and filters to prevent the unwanted noise from disturbing the operation of the device. 
		}
	}
\end{figure*}
\begin{center}
	\bf{SUPPLEMENTARY NOTE 3: Device sketch, system parameters and experimental setup}
\end{center}


The UQCM is demonstrated in a superconducting circuit consisting of five
frequency-tunable Xmon qubits, labeled from $Q_{1}$ to $Q_{5}$, coupled
to a resonator with a fixed frequency of $\omega_{\text{r}}/2\pi = 5.588$ GHz \cite{song_naturecom2017,sup_ning_prl2019}, 
as sketched in Supplementary Figure \ref{cloning_wiring}. $%
Q_{4}$ and $Q_{5}$ (not marked here), not used in the experiment,
stay far off-resonant from the other three qubits and the resonator throughout the copying process. The parameters are detailed in Supplementary Table ~\ref{table1}.
The qubit-resonator coupling strength $g_j$ is measured and estimated through the vacuum Rabi oscillations.
The qubits' energy decaying time $T_1$, the Ramsey dephasing time $T_{2,j}^*$,
and the dephasing time with spin-echo $T_{2,j}^{SE}$ are respectively measured at qubits' idle frequencies $\omega_j$; while the
values in parentheses are those measured at the qubits' working frequency $\omega_{\text{w}}/2\pi=5.44$ GHz where
the resonator-induced qubit-qubit couplings are realized and used for the Bell-state generation and the
cloning operation. The leakage rate $\kappa_{j}^{r}$ of $Q_j$'s readout resonator $R_j$ is obtained through measuring the frequency shift of $Q_j$ induced by photons accumulated in its readout resonator \cite{sank2014,sup_ning_prl2019}. $F_{0,j}$ ($F_{1,j}$) is the probability of detecting $Q_j$ in $\vert 0\rangle$ ($\vert 1\rangle$) when it is prepared in $\vert 0\rangle$ ($\vert 1\rangle$).
More details about the device
performance are described in the Supplementary Materials of Refs. \cite{song_naturecom2017,sup_ning_prl2019}.%

The whole electronics and wiring for the device control is separated into two parts: one at room temperature and the other in the dilution refrigerator, as shown in Supplementary Figure \ref{cloning_wiring}.  Each qubit XY control is implemented through the mixing of the low-frequency signals produced by 2 Digital-to-analog convertor (DAC) channels (I/Q) and a Microwave source (MS) whose carry frequency is $5.52$ GHz; it realizes the fast qubit flipping at the nanosecond scale. Each qubit's frequency control is implemented by the signals from two ways that are combined from a bias tee, one is directly produced by a DAC channel for the fast tuning of the qubit's frequency, while the other is sent by a Direct-current (DC) channel to reset the qubit's frequency to specific operating point. The qubit readout control is achieved also by the sideband mixing of the signals of two independent DAC channels I/Q and a MS with the frequency of $6.67$ GHz, thus it outputs a readout pulse with 3 tones that targets 3 qubits' readout resonators. The output from the sample is amplified sequentially by the impedance-transformed Josephson parametric amplifier (JPA), high electron mobility transistor (HEMT) and room temperature amplifiers before being captured and demodulated by ADC. 4 cryogenic circulators with low insertion loss are placed between JPA and the device to allow the unidirectional-transmission of microwave signal which impedes the reflections and noise from the input entering into the device. The JPA is pumped by a MS with the frequency of $13.494$ GHz, about twice the signal frequency through the pumping line, where the signal is filtered by a $13.1$ GHz band-pass filter; the amplification band of the JPA is tunable with a DC bias applied to it. To reduce kinds of unwanted noises feeding into the signal lines, each signal from (to) DAC and Analog-to-digital convertor (ADC) is filtered with a 7.5 GHz low-pass filter, each qubit Z pulse from DAC is filtered with a $500$ MHz low-pass filter, and each DC signal is filtered with a $80$ MHz low-pass filter at Mixing-Chamber (MC) stage and with a RC filter at 4K state. Moreover, some attenuators are also added to the signal lines at different temperature stages to further reduce the noises influencing  the operation of the device.                                    

\begin{figure*}[!htb]
	\centering
	\includegraphics[width=0.61\textwidth,clip=True]{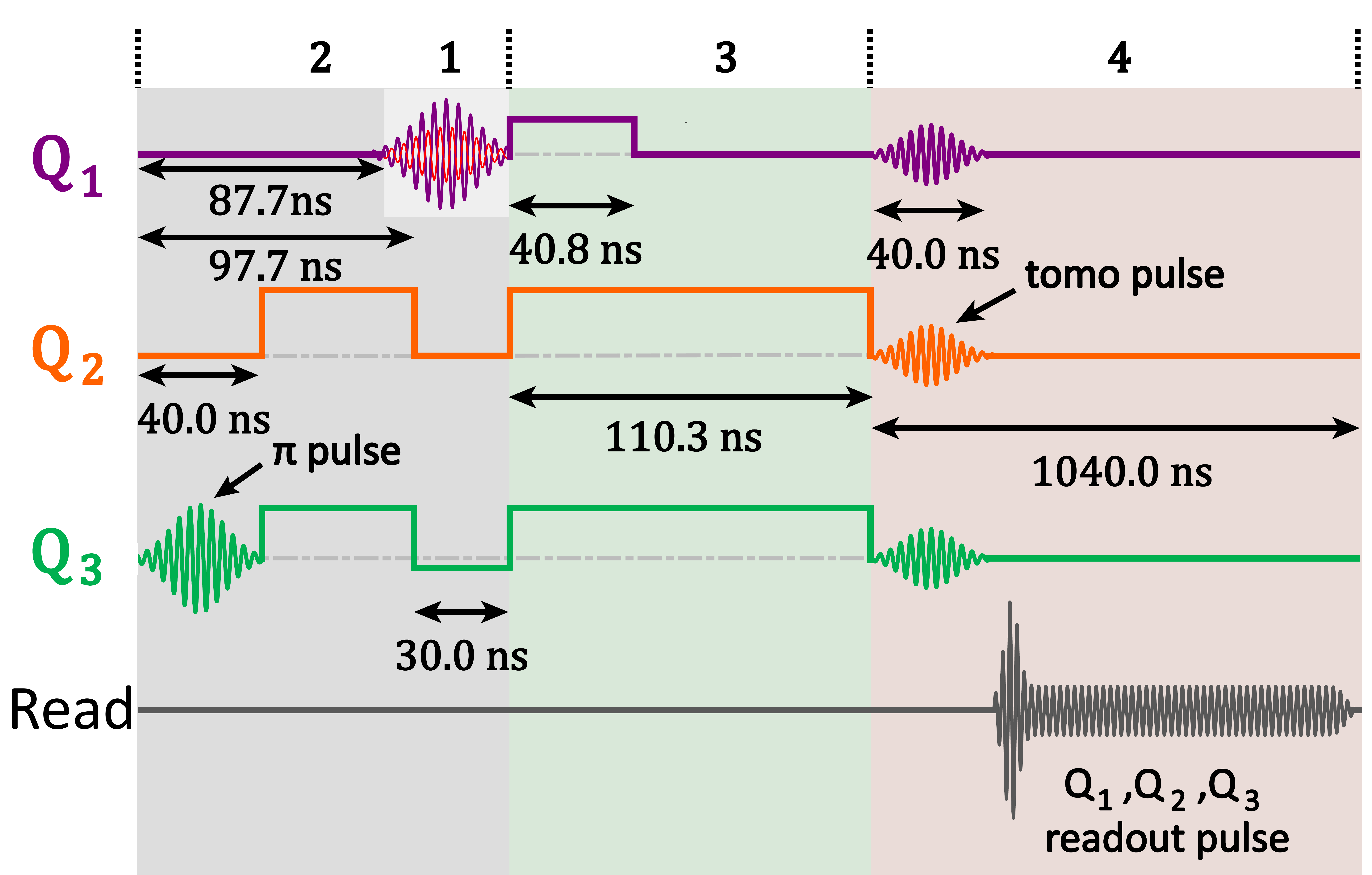}
	\caption{\label{quantumcloning-sequence-v11} \footnotesize{Experimental pulse sequence. The sequence is divided into 3 stages which involves 4 steps: 1. the preparation of the state to be cloned for $Q_1$; 2. the preparation of the entangled state for $Q_2$ and $Q_3$; 3.  resonator-induced couplings $C_{1,2,3}$ and $C_{2,3}$; and 4. quantum state tomography. In the first stage, a microwave $\pi$-pulse with a duration of 40 ns is first applied on $Q_3$ to excite it to $\left|1_3\right\rangle$, then a rectangular pulse with a length of 57.7 ns is applied on the qubit pair $Q_2$-$Q_3$ to tune them on resonance, realizing the $\sqrt{i\text{SWAP}}$ gate which generates the entangled state $\left[e^{i(\pi/2+\theta_{\text{d}})}\left\vert 1_{2}0_{3}\right\rangle +\left\vert 0_{2}1_{3}\right\rangle \right]/\sqrt{2}$. A compensated rectangular pulse of a length of 30 ns is then applied on $Q_3$ to cancel the phase factor $i$ and the dynamical phase $\theta_{\text{d}}$ accumulated during the frequency tuning process. The step 1 for the preparation of the input state to be cloned for $Q_1$ begins in the last 40 ns of the step 2, through a microwave pulse that terminates synchronously with the phase compensated pulse. In step 3, the resonator-induced coupling $C_{1,2,3}$ is started with a rectangular pulse applied on each qubit: $Q_1$, $Q_2$, and $Q_3$, tuning them on resonance for about 40.8 ns; this is successively transformed to the coupling $C_{2,3}$ as the rectangular pulse on $Q_1$ is withdrawn at the moment thus remaining only $Q_2$ and $Q_3$ virtually coupling with the resonator, the duration is about 69.5 ns. Note that a single-qubit z-axis rotation is then numerically applied on each qubit to cancel the extra phase accumulated due to the qubits' frequency shift during the process. The sequence of step 4 involves a tomographic pulse of a duration 40 ns and a readout pulse of a duration of 1 $\mu$s on each qubit. Thus the density matrix for each qubit and the joint density matrices for any two qubits can be reconstructed.}	}
\end{figure*}

\begin{center}
	\bf{SUPPLEMENTARY NOTE 4: Experimental pulse sequences}
\end{center}

The pulse sequence for UQCM is shown in Supplementary Figure \ref{quantumcloning-sequence-v11}, which is divided into 3
stages in time series. The first stage involves 2 steps for the implementation: the preparation of the state 
to be cloned for $Q_1$ and the preparation of the entangled state for the two copy qubits $Q_2$ and $Q_3$. 
The experiment begins with a $\pi$ rotation $X_{\pi}$ applied on $Q_3$, realized by a microwave pulse with a duration of 40 ns at its idle frequency. Then $Q_2$ and $Q_3$ are biased with rectangular pulses from their respective idle frequencies (see Supplementary Table \ref{table1}) to the working frequency $\omega_{\text{w}}/2\pi=5.44$ GHz, which is    
red-detuned from the resonator frequency by an amount of $\Delta=2\pi\times148$ MHz. Thus it implements a $\sqrt{i\text{SWAP}}$ gate and steers $Q_2$-$Q_3$ to the entangled state $\left\vert \psi_{2,3}\right\rangle=\left[e^{i(\pi/2+\theta_{\text{d}})}\left\vert1_20_3\right\rangle+\left\vert0_21_3\right\rangle\right]/\sqrt{2}$, with $\theta_{\text{d}}$ being the dynamical phase accumulated during the frequency tuning process.
A compensated Z-pulse is applied to $Q_3$ for 30 ns to eliminate the factor $e^{i(\pi/2+\theta_{\text{d}})}$ of such an entangled state  $\left\vert \psi_{2,3}\right\rangle$.
Meanwhile, $Q_1$ is prepared to the initial state to be cloned by microwave pulses sent through its XY control line. The second stage launches by biasing each of the three qubits with a rectangular pulse to the working frequency $\omega_{\text{w}}$, realizes the resonator-induced three-qubit coupling $C_{1,2,3}$ for the preparation of the three-body state in Eq. (4) of the main text and the resonator-induced two-qubit coupling $C_{2,3}$ for the neutralization of the phase factor $e^{-i\pi/3}$ in such a three-body state. The processes $C_{1,2,3}$ and $C_{2,3}$ last for 40.8 ns and 69.5 ns, respectively. In the third stage, the qubit output state tomography is performed, where three kinds of tomographic pulses ($I$, $X/2$, $Y/2$) for each qubit are iterated for different experimental run. With this pulse sequence, the density matrices for each of the copying qubits $Q_2$ and $Q_3$ associated with the different input
states $\{\left\vert 0_{1}\right\rangle$, $\left( \left\vert
0_{1}\right\rangle +i\left\vert 1_{1}\right\rangle \right) /\sqrt{2}$, $\left(\left\vert
0_{1}\right\rangle-i\left\vert 1_{1}\right\rangle \right)/\sqrt{2}$, $\left(\left\vert
0_{1}\right\rangle+\left\vert 1_{1}\right\rangle \right)/\sqrt{2}$, $\left(\left\vert
0_{1}\right\rangle-\left\vert 1_{1}\right\rangle \right)/\sqrt{2}$,
$\left\vert 1_{1}\right\rangle \} $ are reconstructed and displayed in Fig. 2a-f of the main text, respectively. While the joint $Q_1$-$Q_2$ and $Q_1$-$Q_3$ output density matrices associated with the six input states are reconstructed and 
displaced in Fig. 4a-f of the main text, respectively.    

\begin{figure*}[!htb]
	\centering
	
	\includegraphics[width=0.71\textwidth,clip=True]{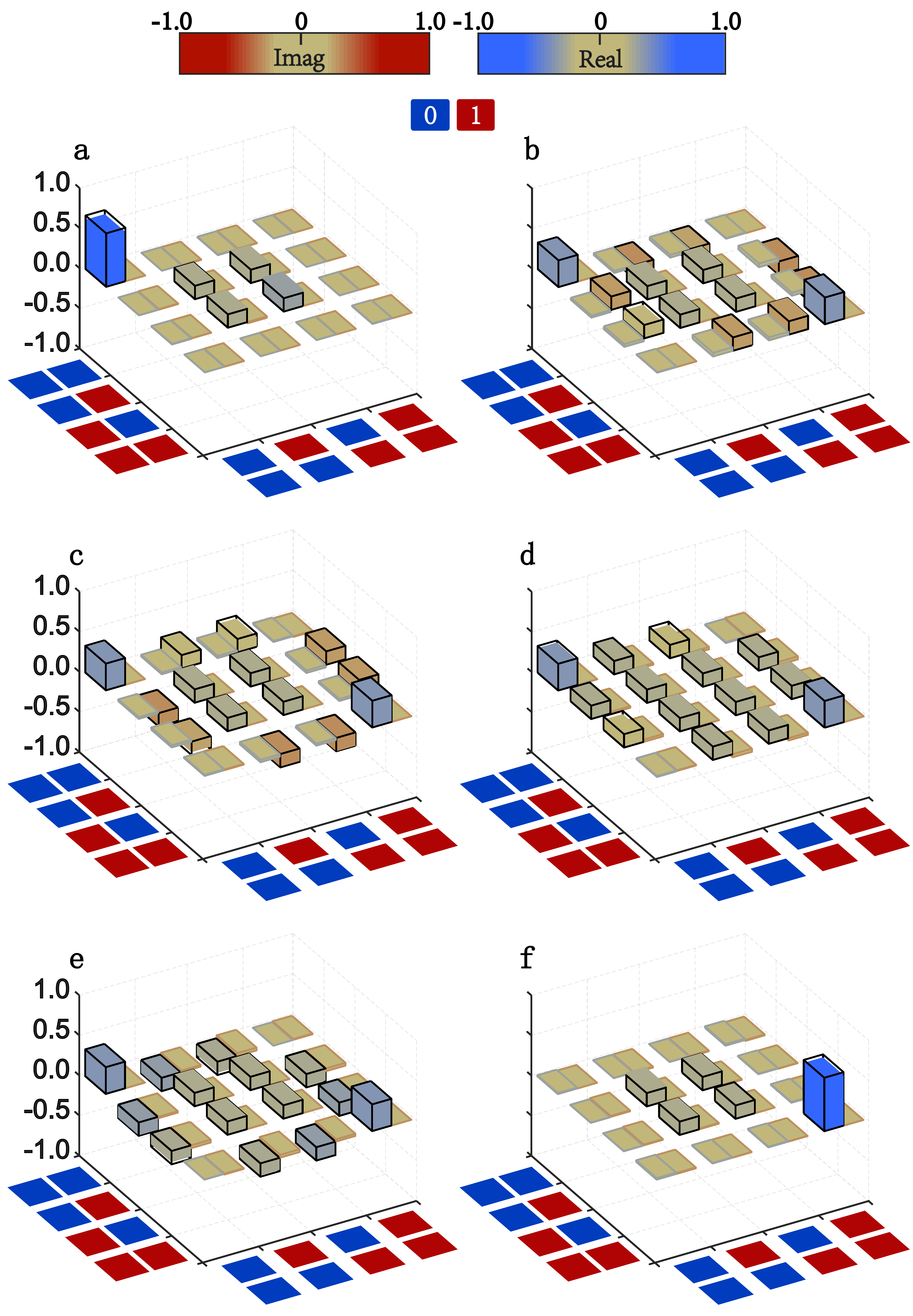}
	\caption{\label{finalstates_V4} \footnotesize{$Q_{2}$-$Q_{3}$ joint output density matrices for the
			six input states: \textbf{a} $\left\vert 0_{1}\right\rangle $; \textbf{b} $\left(
			\left\vert 0_{1}\right\rangle +i\left\vert 1_{1}\right\rangle \right) /\sqrt{%
				2}$;  \textbf{c} $\left(\left\vert 0_{1}\right\rangle -i\left\vert 1_{1}\right\rangle \right) /\sqrt{%
				2}$; \textbf{d} $\left( \left\vert 0_{1}\right\rangle +\left\vert 1_{1}\right\rangle
			\right) /\sqrt{2}$; \textbf{e} $\left( \left\vert 0_{1}\right\rangle -\left\vert 1_{1}\right\rangle
			\right) /\sqrt{2}$; \textbf{f} $\left\vert 1_{1}\right\rangle $. Each matrix element is
			characterized by two color bars, one for the real part and the other for the
			imaginary part. The black wire frames denote the corresponding matrix
			elements of the ideal output states. 
		}
	}
\end{figure*}
\begin{center}\label{sup note 5}
	\bf{SUPPLEMENTARY NOTE 5: $Q_{2}$-$Q_{3}$ joint state tomography}
\end{center}

To characterize the quantum entanglement between the two copy qubits after
the cloning operation, we perform joint output state tomography for $Q_{2}$
and $Q_{3}$. The $Q_{2}$-$Q_{3}$ joint output density matrices for the six
input states $\left\vert 0_{1}\right\rangle$, $\left( \left\vert
0_{1}\right\rangle +i\left\vert 1_{1}\right\rangle \right) /\sqrt{2}$, $\left(\left\vert
0_{1}\right\rangle-i\left\vert 1_{1}\right\rangle \right)/\sqrt{2}$, $\left(\left\vert
0_{1}\right\rangle+\left\vert 1_{1}\right\rangle \right)/\sqrt{2}$, $\left(\left\vert
0_{1}\right\rangle-\left\vert 1_{1}\right\rangle \right)/\sqrt{2}$,
$\left\vert 1_{1}\right\rangle  $ are shown in Supplementary Figure \ref{finalstates_V4}a-f, respectively. The measured matrix elements are denoted by two color bars,
one for the real part and the other for the imaginary part, which are in
well agreement with the corresponding results for a perfect UQCM (black wire
frames). Note that for clarity, before the state tomography,
a single-qubit z-axis rotation with a specific radian angle ($Q1$: -0.390; $Q2$: 0.145; $Q3$: -0.331.) is numerically
applied to cancel the extra phase accumulated due to the
qubits' frequency shift, which does not change the entanglement.

\begin{center}
	\bf{SUPPLEMENTARY NOTE 6: The calibration for the measurement}
\end{center}

In our experiment, the measured probabilities for reconstructing the system states are
calibrated on the basis of the qubits' $\left\vert 0\right\rangle$- and $\left\vert1\right\rangle$-state measurement fidelities, which are shown in Supplementary Table ~\ref{table1}. The calibrated matrix for each qubit is thus constructed to be \cite{song_naturecom2017,sup_ning_prl2019}  
\begin{equation}      
\hat{F}_j^c=\left(                
\begin{array}{cc}   
F_{0,j} & 1-F_{1,j} \\  
1-F_{0,j} & F_{1,j} \\ 
\end{array}
\right).                
\end{equation} 
The system's state density matrix ($\hat{\rho}_N^c$) is reconstructed by correcting the measured probabilities ($\hat{\rho}_N$) by multiplying the matrix inversion as
\begin{equation}
\hat{\rho}_N^c=\hat{F}^{-1} \cdot \hat{\rho}_N,	
\end{equation}
with $\hat{F}$ defined as the tensor product of each qubit's $\hat{F}_j^c$: $\hat{F}= (\hat{F}_j^c)^{\otimes j}$. 

\begin{center}
	\bf{SUPPLEMENTARY NOTE 7: The state and process tomography}
\end{center}
In our experiment, states produced at specific stages are characterized and reconstructed by the technique of quantum state tomography (QST). 
Take a single qubit state, pure or mixed, as an instance, its density operator in the basis $\{|0\rangle, |1\rangle\}$ can be described by 
\begin{eqnarray}\label{es18}
\hat\rho=\rho_{0,0}|0\rangle\langle0|+\rho_{0,1}|0\rangle\langle1|+\rho_{1,0}|1\rangle\langle0|+\rho_{1,1}|1\rangle\langle1|. 
\end{eqnarray}    
Intuitively, such a state can be visually described as a point lying in or on the Bloch sphere. It's then naturally 
to introduce the Bloch vector (connecting the origin of the coordinate to such a point)
\begin{equation}
\vec{r}=\langle \hat{X}\rangle \vec{e}_X + \langle \hat{Y}\rangle \vec{e}_Y + \langle \hat{Z}\rangle \vec{e}_Z,
\end{equation}     
with 
\begin{eqnarray}
\langle \hat{X}\rangle=tr(\hat{\rho}\hat{X})=\rho_{0,1}+\rho_{1,0},\nonumber\\
\langle \hat{Y}\rangle=tr(\hat{\rho}\hat{Y})=i(\rho_{0,1}-\rho_{1,0}),\nonumber\\
\langle \hat{Z}\rangle=tr(\hat{\rho}\hat{Z})=\rho_{0,0}-\rho_{1,1}.
\end{eqnarray}
To locate such a point means to define the state because of $\hat{\rho}=\sum_{\hat{A}=\hat{X},\hat{Y},\hat{Z},\hat{I}}\langle \hat{A}/2\rangle\hat{A}$, which requires the three coordinate components, i.e., measured values $\langle \hat{X}\rangle$, $\langle \hat{Y}\rangle$, and $\langle \hat{Z}\rangle$ along the corresponding axes. For the two-level system, a typical measurement reveals the probability $P_0$ or $P_1$ for it be in state $|0\rangle$ or $|1\rangle$, that directly translates to the component along the Z-axis giving Z-coordinate: $\langle \hat{Z} \rangle=P_0-P_1$, i.e., the diagonal elements in Eq. (\ref{es18}). To get the X-coordinate of the state, we rotate about Y by $-\pi/2$ thus bring the X axis to the original position of the Z axis before measurement; while to get the Y-coordinate, we rotate about X by $\pi/2$ before measurement. The last two measurements after the rotations give the non-diagonal elements of Eq. (\ref{es18}). Therefore these three measurements allow the quantum state $\hat{\rho}$ of Eq. (\ref{es18}) to be completely reconstructed. Suppose a system of N qubits, we perform $3^N$ possible combinations of rotations on the qubits: $U^j\in\{X/2,Y/2,I\}^{{\otimes}N}$, and measure the system to count $2^N$ probabilities of $\{P_{00...0},P_{00...1},...,P_{11...1}\}$: 
\begin{eqnarray}\label{eqs21}
&& P_k^j=\rho_{kk}^j=\langle k|U^j\rho(U^j)^+|k\rangle, \nonumber\\
&& \quad k=1,2,...,2^N, \quad j=1,2,...,3^N.          
\end{eqnarray}
Each combination of rotations needs a pulse sequence, in which we append a tomographic operation for each qubit after producing the state, and then measure the binary outcomes of the N qubits. We repeat the same sequence thousands of times to count all the probabilities. Taking into account the condition of normalization, these measurements produce $3^N(2^N-1)$ numbers to determine the N qubits' density matrix $\hat{\rho_N}$, that possesses $4^N-1$ degrees of freedom. Owing to the condition of $3^N(2^N-1)\ge4^N-1$, the measurement outcomes give more than enough information to determine $\hat{\rho}_N$. By introducing the vector form $\tilde{P}_O=P_k^j$ and $\tilde{\hat{\rho}}_N=\hat{\rho}_N$, we simplify Eq. (\ref{eqs21}) as
\begin{equation}\label{eqs22}
\tilde{P}_O=\tilde{U}\tilde{\hat{\rho}}_N,   
\end{equation} 
with $\tilde{U}$ being the tensor object formed by the combination of $U^j$ and $(U^j)^+$. As the problem in the form of Eq. (\ref{eqs22}) is overconstrained, for our implementation we solved it by turning to a least-squares optimization as to find the optimal solution. Due to the inevitable noises involved in the measured quantities, conversely it is helpful to use the overdetermined equations to guarantee the availability of the outcomes. Nevertheless, the solved $\hat{\rho}_N$ through such a method does not necessarily ensure that it is positive, Hermitian, and simi-definite with unit trace. It is still needed another optimization, that is, $min(\Vert\hat{\rho}_N-\hat{\rho}_N^\prime\Vert)$, to get the required $\hat{\rho}_N^\prime$ that satisfies the three conditions imposed. Then the fidelity of the qubit state can be obtained by the calculation  
\begin{equation}
F=tr(\hat{\rho}_N^\prime*\hat{\rho}_{id}),
\end{equation}             
where $\hat{\rho}_{id}$ is the ideal density matrix. Note that the reconstructed density matrices for the fidelity calculation are calibrated according to the qubits' $\left\vert0\right\rangle$- and $\left\vert1\right\rangle$-state measurement fidelities, as shown in Supplementary note 5.   

The quantum process tomography (QPT) characterizes certain evolution processes for quantum operations. These quantum operations can be described by the operator-sum representation:
\begin{equation}\label{es23}
\mathcal{M}(\hat{\rho})=\sum_i\hat{B}_i\hat{\rho} \hat{B}_i^+.
\end{equation}
Such a linear map $\mathcal{M}$ completely describes the dynamics of the system, whose states could change due to not only any unitary operation performed by kinds of quantum gates, but also projection or decoherence effect that are not unitary. The operators $\hat{B}_i$ act on the system alone, and satisfy the completeness relation $\sum_i \hat{B}_i^+\hat{B}_i=I$ to guarantee $Tr[\mathcal{M}(\rho)]=1$. The key is to determine the operators $\hat{B}_i$ in experiment for a specific quantum process. To do so, it's required to relate the $\hat{B}_i$ to measurable values and expand them with a linear combination of a set of operators $\tilde{B}_i$: 
\begin{equation}
\hat{B}_i=\sum_h b_{ih}\tilde{B}_h,
\end{equation}
while not changing the description of $\mathcal{M}$, provided the $\tilde{B}_h$ form a set of complete operation bases acting on the state space, and $b_{ih}$ are certain complex numbers. The mapping in Eq. (\ref{es23}) is thus rewritten as 
\begin{equation}\label{es25}
\mathcal{M}(\rho)=\sum_{hv}\tilde{B}_h\hat{\rho}\tilde{B}_v^+\chi_{hv}, 
\end{equation}     
where $\chi_{hv}=\sum_ib_{ih}b_{iv}^*$. Obviously, the $\chi_{hv}$ are determined by the $\tilde{B}_h$, and completely describe the whole mapping process provided the $\tilde{B}_h$ are fixed. 

We now show how to determine $\chi_{hv}$ experimentally. Consider a N-qubit system, whose initial state can be expressed as a $2^N \times 2^N $ matrix. The matrix can be written as the unique linear combination of $\hat{\rho}_j$, with $\hat{\rho}_j$ being the linearly independent basis element within the matrix. It's convenient to choose the basis operator $\hat{\rho}_{hv}\equiv |h\rangle\langle v|$, and by preparing a set of preset input states, commonly, such as $|j\rangle \in \{|h\rangle, |v\rangle, (|h\rangle+|v\rangle)/\sqrt{2}, (|h\rangle+i|v\rangle)/\sqrt{2}\}$, after the process transformation due to the system's changing, it is then natural to produce the output $\mathcal{M}(\hat{\rho}_{j})$, which is formed by linear combinations of $\mathcal{M}(|j\rangle\langle j|)$, as can be determined by quantum state tomography. Therefore, it is reasonable to express $\mathcal{M}(\hat{\rho}_j)$ as a linear combination of the basis states: 
\begin{equation}\label{es26}
\mathcal{M}(\hat{\rho}_j)=\sum_k\eta_{jk}\hat{\rho}_k,
\end{equation}
which determines $\eta_{jk}$ based on the known $\mathcal{M}(\hat{\rho}_{j})$. 

Going back to Eq. (\ref{es25}), we write within it the relation:
\begin{equation}
\tilde{B}_h\hat{\rho}_j\tilde{B}_v^+=\sum_k \xi_{jk}^{hv}\hat{\rho}_k,
\end{equation}
which determines $\xi_{jk}^{hv}$ given the $\tilde{B}_h$ operators and the $\hat{\rho}_j$ operators. Combining Eqs. (\ref{es25}) and (\ref{es26}), we have 
\begin{equation}\label{es28}
\sum_k \sum_{hv}\chi_{hv}\xi_{jk}^{hv}\hat{\rho}_k=\sum_k\eta_{jk}\hat{\rho}_k. 
\end{equation}
Eq. (\ref{es28}) immediately gives 
\begin{equation}
\sum_{hv}\chi_{hv}\xi_{jk}^{hv}=\eta_{jk}
\end{equation}
due to that $\hat{\rho}_k$ is linearly independent.The generalized inverse $\varsigma$ of $\xi_{jk}^{hv}$ can be solved with the help of specific software packages for matrix calculation. Thus $\chi_{jk}^{hv}$ can then be obtained to be
\begin{equation}
\chi_{hv}=\sum_{jk}\varsigma_{jk}^{hv}\eta_{jk}. 
\end{equation}

\newpage

%